\documentstyle[aps,preprint]{revtex}

\begin{document}
\draft

\title            {Qualitative viscous cosmology}

\author           {A.A. Coley, R.J. van den Hoogen\footnote{Current 
Address:  Department of Mathematics and Computer Science, St. Francis 
Xavier University, Antigonish, Nova Scotia, B2G 2W5}}

\address          {Department of Mathematics, Statistics, and 
                   Computing Science, Dalhousie University, 
                   Halifax, Nova Scotia, Canada, B3H 3J5}

\author           {R. Maartens  \footnote{Member of Center for 
                  Nonlinear Studies, Witwatersand University, 
                                          South Africa 2050}} 
\address          {School of Mathematical Studies, University of 
                   Portsmouth, Portsmouth, England PO1 2EG}

\maketitle

\begin{abstract}
The Full (non--truncated) Israel--Stewart theory of bulk viscosity is
applied to dissipative FRW spacetimes. Dimensionless variables and
dimensionless equations of state are used to write the
Einstein--thermodynamic equations as a plane autonomous system and the
qualitative behaviour of this system is determined.
 Entropy production in these models is also discussed.
\end{abstract}



\section{Introduction}
 
In a recent paper \cite{Coley95}, isotropic and spatially homogeneous 
viscous
fluid cosmological models were investigated using the Truncated 
Israel--Stewart
\cite{Israel76,Israel79,Israel79b} theory of irreversible 
thermodynamics to
model the bulk viscous pressure.  Although it provides a causal
and stable second order relativistic theory of thermodynamics, the 
Truncated
version of the theory can give rise to very different behaviour than
the Full Israel--Stewart theory 
\cite{Hiscock91,Zakari93,Romano94,Maartens94}.
It can be argued that the Truncated theory
agrees with the Full theory if one uses, instead of the
local equilibrium variables, a generalised temperature and
thermodynamic pressure \cite{Maartens94,Gariel94}.
However, there are difficulties in modelling these generalised 
variables in cosmology. Therefore
the analysis of \cite{Coley95} can only be
regarded as a first step in the study of dissipative processes in the 
universe
utilizing the Full (non--truncated) theory.

For a FRW (Friedmann-Robertson-Walker) cosmology the metric is given by
$$ ds^2=-dt^2+R(t)^2\left[{dr^2\over 1-kr^2}+r^2(d\theta^2+\sin^2
\theta d\phi^2)\right],~~k=0,\pm 1, $$
and the Einstein field equations and the energy conservation equation 
are given by
\begin{eqnarray}
\dot{H}    &=&   -H^2-\frac{1}{6}(3\gamma-2)\rho
                                         -\frac{1}{2}\Pi,   
\label{t dot}\\
\dot\rho   &=&   -3H(\gamma\rho + \Pi) ,                    
\label{p dot}\\
H^2   &=&   \frac{1}{3}\rho-{k\over R^2},                   
\label{fried}
\end{eqnarray}
where $H=\dot{R}/R$ is the Hubble expansion rate (we  restrict 
ourselves to the expanding case only, i.e., $H>0$),
$\rho$ is the
energy density, and the local equilibrium
pressure is assumed to obey $$P=(\gamma-1)\rho,~~1\leq\gamma\leq2,$$
with $\gamma$ constant.

The bulk viscous pressure $\Pi$ obeys the evolution
equation \cite{Hiscock91,Maartens94}
\begin{equation}
\Pi  =    -3\zeta H -\tau\dot\Pi -
\frac{\epsilon}{2}\tau\Pi
\left[3H+\frac{\dot \tau}{\tau}-\frac{\dot \zeta}{\zeta}
-\frac{\dot T}{T}\right],                                           
\label{pi dot}
\end{equation}
where $\zeta\geq 0$ is the bulk viscosity coefficient, $0\leq \tau$ 
($\equiv \zeta\beta_o$ in \cite{Coley95}) is a relaxation coefficient 
for 
transient bulk viscous effects, and $T \geq 0$ is the temperature.
Equation (\ref{pi dot}) with $\epsilon =1$ arises as the simplest way
(linear in $\Pi$) to satisfy
the H-theorem (i.e., for entropy production  to be non--negative
\cite{Maartens94}).
The Truncated theory effectively arises by setting  $\epsilon =0$, 
i.e., it
corresponds to the case where the term in square brackets in equation
 (\ref{pi dot}) is negligible in comparison with the other terms (see 
\cite{Z} for the appropriate conditions). 

The Israel--Stewart theory is derived under the assumption that the
thermodynamical state of the fluid  is close to equilibrium, 
which means that the 
non--equilibrium bulk viscous pressure should be small when compared 
to the
local equilibrium pressure, viz.,
\begin{equation}
|\Pi|<P=(\gamma-1)\rho.
\label{lin}
\end{equation}
If this condition is violated, then one is effectively
assuming that the linear theory holds also in the non--linear regime
far from equilibrium. Such an assumption is unavoidable for viscous
inflationary cosmology \cite{Maartens94}.
For a fluid description of
the matter, equation (\ref{lin}) ought to be satisfied. However, 
note that non-linear viscous effects may arise in a 
phenomenological description of particle creation in the early 
universe \cite{Vereshkov}.

\section{The dynamical system}
 
Equations of state for $\zeta$ and $\tau$ and a temperature law for 
$T$ are needed in order 
for the above system of equations to be closed.
Belinskii et al. 
\cite{Belinskii80} take $\zeta$ and $\tau$ to be
proportional to powers of $\rho$, and this assumption
is extended to $T$
in \cite{Zakari93}. We shall follow
\cite{Coley95} and adopt `dimensionless' equations of state.  That is,
defining the dimensionless density parameter
\begin{equation}
x    =\Omega\equiv  \frac{\rho}{3H^2},                         
\label{x}
\end{equation}
we shall assume that $\zeta/H$ and $\tau H$ are proportional to 
powers of 
$x$, namely,
\begin{equation}
        \frac{\zeta}{H}=3\zeta_ox^m,
          \quad \text{and}\quad 
        \frac{\tau^{-1}}{H}=bx^n,                             
\label{eqs state}
\end{equation}
where $m$ and $n$ are constants which are assumed to be non-negative 
and 
$\zeta_o$ and $b$ are positive parameters.
Clearly the equations of state employed will determine the qualitative 
properties of the models
\cite{Coley95,Hiscock91,Zakari93,Romano94,Maartens94,Belinskii80}.
Equations of state (\ref{eqs state}), which ensure that the asymptotic
limit points represent self-similar models \cite{Coley94b}, are
phenomological in nature and are no less appropriate than the
equations of state used by Belinskii et al. \cite{Belinskii80}.
We note that the equations of state chosen in
\cite{Belinskii80} and those above coincide in the important case
$m=1/2= n$
($q=1/2$ in \cite{Maartens94}).

From now on we shall take $$n=0,~~a\equiv b\zeta_o.$$ (Note that $a,b$ 
are 
precisely the parameters used in \cite{Coley95}.)
When $n=0$, it follows that
the relaxation rate is determined by the expansion rate:
\begin{equation}
\tau^{-1}=bH.
\label{rate}
\end{equation}
As argued in \cite{Maartens94}, for viscous expansion to be 
non--thermalising, we should have $\tau^{-1}<H$, for otherwise the
basic interaction rate for viscous effects could be sufficiently rapid 
to restore equilibrium as the fluid expands. Therefore we impose the
constraint $$b<1$$ on the relaxation parameter.

Defining the dimensionless viscous pressure $y$ and the new time 
variable $\bar t$ by
\begin{equation}
         y  =       \frac{\Pi}{H^2},
            \quad \text{and}\quad 
   \frac{{\rm d}\bar t}{{\rm d} t} = H,                \label{y}
\end{equation}
and using equation (\ref{t dot}), equations (\ref{p dot}) and 
(\ref{pi dot}) become:
\begin{eqnarray}
x'    &=& (x-1)[(3\gamma-2)x+y],                                  
\label{x'}\\
y'    &=& y[2-b+y+(3\gamma-2)x]-9ax^m-\frac{\epsilon}{2}y\Psi,   
\label{y'}
\end{eqnarray}
where
\begin{equation}
\Psi  \equiv 3-2\frac{H'}{H}-m\frac{x'}{x}-\frac{T'}{T}, \label{Psi}
\end{equation}
and $'$ denotes a derivative with respect to $\bar t$.
Note that the linear condition (\ref{lin}) becomes
$$
|y|<3(\gamma-1)x.
$$

Equations (\ref{x'}) and (\ref{y'}) constitute a plane autonomous 
system of
ODEs for $x$ and $y$.  In the Truncated theory $\epsilon =0$, whence 
the final term in 
equation (\ref{y'}) is absent and there is no need to specify an 
equation for
$T$.  Hereafter we shall set $\epsilon=1$, and adopt  the following 
  temperature power-law \cite{Zakari93,Romano94,Maartens94}:    
\begin{equation}
T=T_0 \rho^r=T_03^rx^{r} H^{2r},\quad{\rm with }\quad 
r=\frac{\gamma-1}{\gamma}
\label{T}
\end{equation}
where the form of the exponent $r$ follows from the integrability 
condition of the Gibbs equation when $P=(\gamma-1)\rho$ \cite{Z,M}.  
When 
the local equilibrium state of the expanding viscous fluid is 
thermalized 
radiation, then $r=1/4$, in line with the
 standard Stefan--Boltzmann relation.  Consequently 
\begin{eqnarray}
\Psi&=&3-2(1+r)\frac{H'}{H}-(m+r)\frac{x'}{x},\nonumber\\
    &=& c_0 + c_1y+c_2x+c_3\frac{y}{x},\nonumber
\end{eqnarray}
where
\begin{eqnarray}
c_0&=& 5+2r+(3\gamma-2)(m+r),\nonumber\\
c_1&=& 1-m,\nonumber\\
c_2&=& (3\gamma-2)c_1,\nonumber\\
c_3&=& m+r.
\nonumber
\end{eqnarray}

\subsection{Flat universe}

All of the FRW models are governed by equations (\ref{x'}) and 
(\ref{y'}) together with equation (\ref{fried}).  We note from 
(\ref{x'}) that $x=1$ is an invariant set, where from
(\ref{fried}) we see that this set represents the flat FRW models.  
Let us study this physically important zero-curvature case first.  
When $x=1$, the thermodynamic laws are simplified to (\ref{rate}),
(\ref{T}) and, by (\ref{eqs state}), to
\begin{equation}
\zeta \propto H.
\label{bv}
\end{equation}
Thus the bulk viscosity coefficient, like the relaxation rate, is also
determined by the expansion rate. Furthermore,
\begin{equation}
\Psi=(c_0+c_2)+(c_1+c_3)y,\nonumber
\end{equation}
whence equation (\ref{y'}) becomes
\begin{equation}
y'=-\frac{(r-1)}{2}y^2-by-9a.
\label{13}
\end{equation}
That is, the equations governing the evolution of the flat FRW viscous 
fluid models reduce to a single autonomous ODE in $y$.  
Since $0\leq r\leq 1/2$,  (\ref{13}) is a Riccati equation with 
constant 
coefficients and its solutions can be found in implicit form.  
 
Defining the positive parameter 
$$B_1\equiv b^2+18a(1-r),$$ 
it follows that  there are two equilibrium points, one positive, one 
negative, 
\begin{equation}
y^\pm = \gamma(b \pm\sqrt{B_1}),
\label{special}
\end{equation}
where one is a sink and the other is a source
(with respect to the invariant set $x=1$, not the full set of all FRW 
models). 
The points (\ref{special}) correspond to the special solutions found
in \cite{Maartens94b} and re--discovered in \cite{Banerjee}.
 
Therefore, the behaviour of the flat models  
using the Full (non-truncated) theory is qualitatively the same as 
the  behaviour in the Truncated theory \cite{Coley95}. 
Of course, this qualitative similarity only holds for the  
restrictive thermodynamic laws (\ref{rate}), (\ref{T}) and (\ref{bv}).

\subsection{Curved universes}

Let us now return to the general curvature case $x\not =1$ [see  
equations
(\ref{x'}) and (\ref{y'})].  Equation (\ref{y'}) can be written as
\begin{equation}
y'=-y\biggl[\left(b-2+\frac{c_0}{2}\right) +
       x(3\gamma-2)\left(\frac{c_1}{2}-1\right) + 
y\left(\frac{c_1}{2}-1\right)
+ \frac{c_3}{2}yx^{-1}\biggr]-9ax^m.\label{14}
\end{equation}
There are two equilibrium points lying in the invariant set $x=1$, 
namely 
$(1,y^\pm)$ where $y^\pm$ is given by equation (\ref{special}).  
Previously (in the 
case of the flat models) we 
considered the stability of the equilibrium points only with respect 
to the 
invariant set $x=1$; let us now discuss the stability of these 
equilibrium 
points with respect to the curved FRW models.  The equilibrium point 
$(1,y^+)$ is a source with the invariant set $x=1$ as one of its 
primary 
eigendirections.  If $y^-+3\gamma>2$ then the equilibrium point 
$(1,y^-)$ is 
a saddle with $x=1$ as the stable manifold.  
If $y^-+3\gamma<2$ then the
 equilibrium point $(1,y^-)$ is a sink and hence it represents a 
 future asymptotic attractor.
From equation (\ref{x'}), the equilibrium points $(\bar x, \bar y)$ 
not 
lying in the invariant set $x=1$ satisfy $\bar y=-(3\gamma-2)\bar x$, 
and hence from (\ref{14}) we obtain
\begin{equation}
        9a{\bar x}^m-\frac{1}{2}(3\gamma-2)(2b+2r+1)\bar x=0.       
\label{15}
\end{equation}
For $m>0$, there exists a singular point at the origin $(0, 0)$.  
[Note, 
however, that the system of ODEs as given by equations
(\ref{x'}) and (\ref{14}) is not defined at $x=0$ except when 
$c_3=0$ and therefore the point $(0,0)$ may not be a well defined
 equilibrium point of the system.]
 Changing to polar coordinates, it can be shown  that this singular 
 point 
is saddle-like in nature (hyperbolic sectors) if $m<1$.   
If $m>1$, then 
the point $(0,0)$
has  parabolic and hyperbolic sectors.

If $m\not = 1$, then there is  a second equilibrium point  at
\begin{equation}
(\bar x,\bar 
y)=\Biggl(\left[\frac{(3\gamma-2)}{18a}(2b+2r+1)\right]^{1/(m-1)},  
-(3\gamma-2)\bar x \Biggr).\nonumber
\end{equation}
If $$B_2\equiv (3\gamma-2)(2b+2r+1)-18a>0,$$ then $\bar x>1$, and 
when $m<1$, 
this point is a saddle.  If $B_2<0$, then $\bar x<1$, and when $m>1$, 
this equilibrium point is again a saddle.   There is a 
variety of other possible behaviours.

\section{Discussion}

\subsection{Exact Solutions and Asymptotic Behaviours}

The qualitative behaviour of the flat FRW models has been determined 
completely.
 The unphysical flat models evolve from the 
equilibrium point $y=y^+$ at $\bar t=-\infty$, where $y=y^+$ 
corresponds 
to the solution  (after  recoordination)
\begin{eqnarray}
&& R(t)=R_0t^{2/(y^++3\gamma)},\qquad\qquad 
H(t)=\frac{2}{y^++3\gamma}t^{-1},
\nonumber\\
&& \rho(t)=\frac{12}{(y^++3\gamma)^2}t^{-2},\qquad\qquad
\Pi(t)=\frac{4y^+}{(y^++3\gamma)^2}t^{-2},\label{sol1}
\end{eqnarray}
towards either  points at infinity or to the point $y=y^-$ (at $\bar 
t=-\infty$),
which, if $y^-\not = -3\gamma$, has solution
\begin{eqnarray}
&& R(t)=R_0(t-t_0)^{2/(y^- +3\gamma)},\qquad\qquad 
H(t)=\frac{2}{y^-+3\gamma}(t-t_0)^{-1}
\nonumber\\
&& \rho(t)=\frac{12}{(y^-+3\gamma)^2}(t-t_0)^{-2},\qquad\qquad
\Pi(t)=\frac{4y^-}{(y^-+3\gamma)^2}(t-t_0)^{-2}.\label{sol2}
\end{eqnarray}
[Note that if $y^-+3\gamma>0$, then the solution (\ref{sol2}) can be 
recoordinatized such that $t_0=0$.]
These models and the equilibrium point $y=y^+$ are unphysical since
they have positive bulk viscous pressure. Those models which evolve
towards $y=y^-$ have negative bulk viscous pressure after a certain
time, and may be considered as physical models after this time. The
models which are physical for all times (i.e. which have $\Pi<0$ for
all times) are (\ref{sol2}) and those which evolve from infinity at 
$\bar{t}=-\infty$ towards $y=y^-$ at $\bar{t}=\infty$."
[Note that, by (\ref{y}), (\ref{sol1}) and (\ref{sol2}), 
$\bar{t}=-\infty$ corresponds to
$t=0$, while $\bar{t}=\infty$ corresponds to $t=\infty$.]

If $y^-=-3\gamma$ then the solution has the form
\begin{eqnarray}
&& R(t)=R_0{\rm e}^{H_0t},\qquad\qquad H(t)=H_0\nonumber\\
&& \rho(t)=3H_0^{\,2},\qquad\qquad
\Pi(t)=y^-H_0^{\,2}.\label{sol3}
\end{eqnarray}

The exponential inflationary solution (\ref{sol3}) clearly violates 
the condition  
(\ref{lin}) (cf. \cite{Maartens94}). 
The solution (\ref{sol2}) violates (\ref{lin}) if $y^-+3\gamma<3$, 
when the expansion is driven by a large and effective 
nonlinear bulk viscous pressure.
  The expansion 
is from a big bang, and is more rapid than in the corresponding
equilibrium solution
($y^-=0$). Indeed, if 
\begin{equation}
y^-+3\gamma<2,\label{lin2}
\end{equation}
 then the solution represents a power--law inflationary solution.
Knowing that condition (\ref{lin2}) is also the requirement that the 
equilibrium 
point $(1,y^-)$ be stable, we can conclude that the power-law 
inflationary
solution (\ref{sol3}) is the future asymptotic attractor for all 
bulk-viscous 
inflationary FRW models.

We emphasise that bulk viscous inflationary solutions, such 
as (\ref{sol2}),
violate the condition (\ref{lin}), so that their existence is 
dependent on 
assuming
that the theory holds in the nonlinear regime. Furthermore, these
inflationary solutions are limited by the simple equation of
state $P=(\gamma-1)\rho$, so that, in particular, they cannot account
for the processes necessary to provide an exit from inflation. 
The solutions
are at most valid during inflation, and more realistic models would be
needed to incorporate exit and re-heating.

\subsection{Entropy Production}

On physical grounds, one expects that $y\leq 0$, since the evolution
of specific entropy is given by \cite{Hiscock91,Maartens94}
\begin{equation}
\dot{s}=-{3H\Pi\over nT},
\label{sdot}
\end{equation}
where $n$ is the number density. We  note that solution
(\ref{sol2}) always satisfies $y\leq0$.
  From equation (\ref{sdot}), it follows that the growth of entropy
in a comoving volume between times $t_0<t<t_1$ is
given by
\begin{equation}
\Sigma(t_1)-\Sigma(t_0)=-\frac{3}{k}\int_{t_0}^{t_1}
\frac{\Pi H R^3}{T} {\rm d} t,
\end{equation}
where $k$ is the Boltzmann constant.  The amount of entropy generated 
can be 
calculated for each of the solutions (\ref{sol1}), (\ref{sol2}), and 
(\ref{sol3}).  Analyzing the physically more relevant case 
(\ref{sol2}), we
find that (re-instating constants previously set to unity),
\begin{equation}
\Sigma(t_1)-\Sigma(t_0)=\frac{\gamma 3^{(1-r)}c^2}{8\pi G 
k}\left(\frac{R_0^{\ 3}}{T_0}\right)
\left(\frac{2}{y^-+3\gamma}\right)^{2(1-r )} \left(t_1^{\ 
-2y^-/[\gamma(y^-+3\gamma)]}-t_0^{\ 
-2y^-/[\gamma(y^-+3\gamma)]}\right),\label{entropy prod}
\end{equation}
where $c$ is the speed of light and $G$ is the gravitational constant. 
By equation (\ref{sol2}), we must have    $y^-+3\gamma>0$ for an 
expanding solution.
This ensures that (\ref{entropy prod}) gives 
$\Sigma(t_1)>\Sigma(t_0)$.

The amount of entropy generated in the de Sitter model (solution 
(\ref{sol3})) is the same as that calculated by Maartens 
\cite{Maartens94}, namely
\begin{equation}
\Sigma(t_1)-\Sigma(t_0)=\frac{3^{(1-r)}c^2}{8\pi G k}\left(
\frac{R_0^{\ 
3}}{T_0}\right) H_0^{\ 2-2r}\left({\rm e}^{3H_0t_1}-{\rm 
e}^{3H_0t_0}\right).
\end{equation}
It is shown that bulk viscous inflation can generate
significant amounts of entropy without re-heating. 
For the de Sitter model, using
typical parameters of inflation, and assuming that almost all of the
entropy is
produced by inflation, one finds 
the following value for the amount of entropy produced during 
exponential inflation
 \cite{Maartens94}
\begin{equation}
\Sigma \approx 2.1 \times 10^{87},
\end{equation}
which is in agreement with the expected value.
  The power-law inflationary solution (\ref{sol2})
  [with $y^-+3\gamma<2$, i.e., satisfying equation (\ref{lin2})] has 
less efficient entropy production, 
 but nonetheless can also produce significant amounts 
of entropy.

In the above,
we have only considered entropy production in the models corresponding
to the equilibrium points
of the dynamical system.  By considering a simple example, we can
investigate the entropy production in the more general flat FRW models.
 We choose parameter values $r=1/4$ (necessarily $\gamma=4/3$), 
 $a=1/27$, and $b=1/4$.
In this case the differential equation (\ref{13}) reduces to
\begin{equation}
y'=\frac{3}{8}\left[(y-\frac{1}{3})^2-1\right],
\end{equation} 
which has a solution of the form (neglecting the constants of 
integration)
\begin{equation}
y = \left\{
      {\frac{1}{3}-\tanh(\frac{3}{8}\bar t)    \qquad  
      |y-\frac{1}{3}|<1}
      \atop
      {\frac{1}{3}-\coth(\frac{3}{8}\bar t)    \qquad  
      |y-\frac{1}{3}|>1} 
    \right.
\end{equation}
The Hubble parameter is given by
\begin{equation}
H=\left\{ 
 {{\rm e}^{-\frac{13}{6}\bar t}\cosh^{4/3}(\frac{3}{8}\bar t)\qquad 
 |y-\frac{1}{3}|<1}
\atop
{{\rm e}^{-\frac{13}{6}\bar t}\sinh^{4/3}(\frac{3}{8}\bar t)\qquad 
 |y-\frac{1}{3}|>1} 
 \right.
\end{equation}
and $R=R_0{\rm e}^{\bar t}$.
The change in entropy in a comoving volume produced
between times $\bar t_0 < \bar t <
\bar t_1$ is then given by
\begin{equation}
\Sigma(\bar t_1)-\Sigma(\bar t_0)= 
\left(\frac{3^{-1/4}c^2R_0^{\ 3}}{16\pi G k T_0}\right)\left[ 
\pm 2({\rm e}^{-\bar t_1/4}-{\rm e}^{-\bar t_0/4}) +({\rm e}^{-\bar 
t_1}-{\rm e}^{-\bar t_0})
+({\rm e}^{\bar t_1/2}-{\rm e}^{\bar t_0/2})\right].
\end{equation}
It can be concluded in this simple model for $\bar t_1/\bar t_0 >1$ 
that as $\bar t_1$ 
increases the entropy
in a comoving volume grows exponentially with respect to $\bar t_1$.

\section{Conclusions}

The behaviour of the viscous fluid FRW models where the bulk viscous
pressure satisfies the Full Israel-Stewart theory of irreversible
thermodynamics has been analyzed.  
The stability of the equilibrium point $(0,0)$ representing the
Milne model depends upon the value of $m$ which appears in the 
equation
of state for the bulk viscosity.    The  equilibrium point
$(\bar x, \bar y)$ can represent either an open, flat or
closed FRW model depending upon the value of the parameter $B_2$. 
Exact determination of the nature of 
this particular equilibrium point is extremely difficult.
However, a partial result is possible: if $B_2(1-m)>0$, then
the equilibrium point is a saddle.    There exist two equilibrium 
points with
qualitative behaviour similar to that found using the
Truncated Israel-Stewart theory.

It can be concluded that the behaviour of the FRW models
in which the bulk viscous pressure satisfies the Full Israel-Stewart
theory can in principle be qualitatively 
similar to the behaviour of the FRW models in the Truncated theory.  
  One cannot say, however, that the Full theory has the same behaviour 
as the Truncated theory in all cases because it is not at all clear 
what 
effects the presence of anisotropies or different equations of state 
will 
have.  For example, in the models studied here, it was the equations 
of 
state for the temperature and for the bulk viscosity coefficient that 
played 
major roles in determining the dynamics of the models. In the case of 
a 
relativistic Maxwell-Boltzmann gas, which has very
different equations of state, the Truncated and Full theories can lead
to very different behavior, with the Truncated theory leading to
pathological behavior of the temperature in many cases 
\cite{Hiscock91}.

As stated above, the consistency condition that viscous
expansion should be non--thermalising requires $b<1$. 
Further constraints may arise from 
entropy arguments. The evolution equation (\ref{pi dot}) already 
guarantees that entropy production is non--negative. But one may place
constraints on the rate and amount of entropy production. If we impose 
the 
requirement that the specific entropy, $s$, should increase with 
expansion, 
but at a decreasing rate, then we have $y<0$ and possibly further 
constraints
on   $r$ (equivalently $\gamma$) and $m$.

A complete analysis of the asymptotic behaviours of these viscous 
fluid models
depending on the 
(many) free parameters in the model $(a,b,\gamma,m)$ and utilizing the
energy conditions can be made. 
However, the next step in this research 
programme is to attempt to use results from kinetic theory in 
order to motivate physically plausible equations of state, or, at the 
very least, to limit the form of the phenomological equations of 
state used.

\acknowledgements
This work was supported by  
research grants from the Natural 
Sciences and Engineering Research Council of Canada (AAC) and 
Portsmouth University (RM) and through a Killam scholarship (RVDH).



\begin{thebibliography}{10}

\bibitem{Coley95}
A.~A. Coley and R.~J. van~den Hoogen, Class. Quantum Grav. {\bf 12}, 
1977 (1995).

\bibitem{Israel76}
W. Israel, Ann. Phys. {\bf 100},  310  (1976).

\bibitem{Israel79}
W. Israel and J.~M. Stewart, Proc. R. Soc. Lond. A. {\bf 365},  43  
(1979).

\bibitem{Israel79b}
W. Israel and J.~M. Stewart, Ann. Phys. {\bf 118},  341  (1979).

\bibitem{Hiscock91}
W.~A. Hiscock and J. Salmonson, Phys. Rev. D {\bf 43},  3249  (1991).

\bibitem{Zakari93}
M. Zakari and D. Jou, Phys. Rev. D {\bf 48},  1597  (1993).

\bibitem{Romano94}
V. Romano and D. Pavon, Phys. Rev. D {\bf 50}, 2572 (1994).

\bibitem{Maartens94}
R. Maartens, Class. Quantum Grav. {\bf 12}, 1455 (1995).

\bibitem{Gariel94}
J. Gariel and G. le Denmat, Phys. Rev. D {\bf 50}, 2560 (1994).

\bibitem{Z} W. Zimdahl, preprint astro-ph 9601189.

\bibitem{Vereshkov}
G.~M. Vereshkov, Yu.~S. Grishkan, S.~V. Ivanov, V.~A. Nesterenko, and 
A.~N. Poltavtsev, 
Sov. Phys.  JETP. {\bf 46}, 1041 (1977).

\bibitem{Belinskii80}
V.~A. Belinskii, E.~S. Nikomarov, and I.~M. Khalatnikov, Sov. Phys. 
JETP {\bf  50},  213  (1979).

\bibitem{Coley95b}
R.~J. van~den Hoogen and A.~A. Coley, Class. Quantum Grav. {\bf 12},
 2335  (1995).

\bibitem{Coley94b}
A.~A. Coley and R.~J. van~den Hoogen,  in {\em {D}eterministic {C}haos 
in {G}eneral {R}elativity {\rm edited by D. Hobill, A. Burd, and A. 
Coley}}
  ({N}{A}{T}{O} {A}{S}{I} 332{B}, {P}lenum, {N}ew {Y}ork, 1994).

\bibitem{M} R. Maartens preprint, Portsmouth University (1995).

\bibitem{Maartens94b}
R. Maartens and A. M. Kgathi, unpublished (1994).

\bibitem{Banerjee}
N. Banerjee and A. Beesham, preprint (1995).

 

\end{thebibliography}
\end{document}